\documentclass[sn-mathphys,Numbered]{sn-jnl}% Math and Physical Sciences Reference Style
\usepackage{graphicx}%
\usepackage{multirow}%
\usepackage{amsmath,amssymb,amsfonts}%
\usepackage{amsthm}%
\usepackage{mathrsfs}%
\usepackage[title]{appendix}%
\usepackage{xcolor}%
\usepackage{textcomp}%
\usepackage{manyfoot}%
\usepackage{booktabs}%
\usepackage{algorithm}%
\usepackage{algorithmicx}%
\usepackage{algpseudocode}%
\usepackage{listings}%
\usepackage{caption}
\usepackage{subcaption}
\usepackage[usestackEOL]{stackengine}
\theoremstyle{thmstyleone}%
\theoremstyle{thmstyletwo}%

\theoremstyle{thmstylethree}%

\begin{document}


\title{Dissecting Medical Referral Mechanisms in Health Services: Role of Physician  Professional Networks}
%\author{Anonymous submission}

\iffalse
\author{\name Regina de Brito Duarte \email regina\_duarte@outlook.pt \\
       \addr Instituto Superior Técnico\\
       University of Lisbon\\
       Lisbon, 1049-001, Portugal
       \AND
       \name Qiwei Han\email qiwei.han@novasbe.pt \\
       \addr Nova School of Business and Economics\\
       Universidade Nova de Lisboa\\
       Carcavelos, 2755-405, Portugal
       \AND
       \name Claudia Soares\email claudia.soares@fct.unl.pt \\
       \addr Nova School of Science and Technology\\
       Universidade Nova de Lisboa\\
       Caparica, 2825-149, Portugal}
\fi

\author[1]{\fnm{Regina} \sur{de Brito Duarte}}

\author[2]{\fnm{Qiwei} \sur{Han}}

\author*[3]{\fnm{Claudia} \sur{Soares}}\email{claudia.soares@fct.unl.pt}

\affil[1]{\orgdiv{Instituto Superior Técnico}, \orgname{University of Lisbon}, \orgaddress{\city{Lisbon}, \postcode{1049-001}, \country{Portugal}}}

\affil[2]{\orgdiv{Nova School of Business and Economics}, \orgname{Universidade Nova de Lisboa}, \city{Carcavelos}, \postcode{2755-405}, \country{Portugal}}

\affil*[3]{\orgdiv{Nova School of Science and Technology}, \orgname{Universidade Nova de Lisboa}, \orgaddress{\city{Caparica}, \postcode{2825-149}, \country{Portugal}}}

\abstract{
Medical referrals between primary care physicians (PC) and specialist care (SC) physicians profoundly impact patient care regarding quality, satisfaction, and cost. This paper investigates the influence of professional networks among medical doctors on referring patients from PC to SC. Using five-year consultation data from a Portuguese private health provider, we conducted exploratory data analysis and  constructed both professional and referral networks among physicians. We then apply Graph Neural Network (GNN) models to learn latent representations of the referral network. Our analysis supports the hypothesis that doctors' professional social connections can predict medical referrals, potentially enhancing collaboration within organizations and improving healthcare services.  This research contributes to dissecting the  underlying mechanisms in primary-specialty referrals, thereby providing valuable insights for enhancing patient care and effective healthcare management.
}


\maketitle

%\keywords{Keyword1, Keyword2, Keyword3}

%Medical referrals between primary care doctors and specialists affect many aspects of patient care, such as quality of care, patient satisfaction, and health care costs. In this work, we hypothesize that the professional network of medical doctors can influence the referral process. We analyze primary-specialty referrals through transaction data over medical appointments gathered between 2012 and 2017 in all hospital centers of a Portuguese private health provider. The study's main objective is to discover patterns and hidden mechanisms in primary-specialty referrals to improve the efficiency of healthcare services. First, we conduct exploratory data analysis and identify patterns that may affect the formation of doctors' professional networks. Then, we create and analyze the doctor's professional and referral networks. We learn the representation of the referral network using a Graph Neural Network (GNN). Finally, we discuss the node's representation of the referral network, which supports the hypothesis that the professional social connections of doctors are predictive of medical referrals. This work addresses the discovery of important patterns in medical referrals to, in future work, improve the efficiency of collaboration within organizations.

%\begin{document}

%\flushbottom
%\maketitle
\thispagestyle{empty}

\section{Introduction}
Medical referrals between primary care physicians (PC) and specialist care (SC) physicians are an integral part of the healthcare system. These referrals can significantly impact patient care, influencing quality, satisfaction, and cost\cite{Glenn1987,Forrest2006, Barnett2012}. The process entails the transfer of patient care from the PC to the SC, who possesses specialized expertise and resources to manage the patient's condition\cite{shortell1971physician}. Yet, this process is not without challenges such as coordination and communication issues, potential delays, duplication of tests and treatments, and fragmentation of care. For instance, communication breakdowns during referral can lead to delayed diagnoses or even incorrect treatments.

The central hypothesis of this research paper is that professional networks among medical physicians significantly influence the mechanisms of medical referrals. This hypothesis has been informed by previous qualitative studies that have identified how personal relationships between physicians could affect the referral process\cite {muzzin1992understanding,chan2013understanding}. These studies have highlighted how physicians' relationships, both cooperative and conflicting, can dictate referral decisions. Other research has examined various models of collaboration between PC and SC, and has identified the competencies that promote effective collaboration in patient care\cite{berendsen2007motives}. Furthermore, it has been shown that the handover between primary and secondary care can significantly impact patient health outcomes\cite{ekwegh2020improving}. To this end, data from insurance claims has been utilized to construct a patient-sharing network between physicians based on the shared patients they treat, which provides an operationalization of an informal information-sharing network\cite{appel2018social, Barnett2012}. However, this network may not necessarily reflect the formal organizational structure with which physicians are affiliated. Thus, this research aims to uncover hidden mechanisms in primary-specialty referrals by leveraging features extracted from physicians' professional networks. We do so by examining a large-scale dataset collected over five years from a Portuguese health organization. The dataset comprises 12 million consultations between 1.4 million patients and 3,632 physicians and involves a wide range of PC and SC. The analysis involves creating two distinct networks: 1) a referral network, where a connection between a PC and an SC is established if a patient consults a PC and then an SC within a month, and 2) a professional network of physicians connected through shared professional profiles. We selected Graph Neural Network (GNN) models to model the referral network because they excel in learning latent representations of network structures\cite{hamilton2017inductive}. GNNs' effectiveness has been demonstrated in various applications such as protein-ligand binding affinities\cite{son2021development}, protein interaction prediction\cite{xiao2020graph}, rumor detection in social media\cite{choi2021dynamic}, and outlier detection on multivariate weather time-series\cite{li2021dynamic}.

Our research contributes to understanding the impact of physicians' professional networks on the referral process and how network structure from professional networks improves the predictive accuracy of a referral recommendation model\cite{An2018a,almansoori2012electronic}. The overarching aim is to enhance the quality and efficiency of healthcare services and patient health outcomes, which may have practical implications for improving the referral process through targeted recommendations\cite{han2018}.

\section{Related Work}

\subsection{Social Networks' Influence on Medical Referrals}
Numerous studies in the context of US health systems show that specialty referrals are affected by a myriad of factors, including patient, physician, and healthcare system structural characteristics~\cite{Forrest2006}. In particular, specialty referral decision-making is more likely to happen within PC' professional networks~\cite{Barnett2011}, e.g., referrals to SC with whom PC has had prior experiences~\cite{Kinchen2004} or who share similar characteristics~\cite{Landon2012}. Patients also acquire health information and physician recommendations through online health communities~\cite{Yao2021}.

Specialty referrals have increased substantially in the US~\cite{Barnett2012a}, and there are significant variations in PC' referral rates~\cite{Frank2000}, as well as the cost and intensity of care in hospitals~\cite{Barnett2012b}. This can pose several challenges for health organizations to manage the medical referral process efficiently:
\begin{enumerate}
   \item  Closer care coordination between PC and SC is a critical success factor in value-based care. Still, communication gaps between them may persist without organizational facilitation, leading to negative consequences for patients~\cite{Malley2011,Zuchowski2014}.
   \item  The social nature of specialty referrals can lead to substantial variations in referrals received by SC, which can cause organizational inefficiency, e.g., for new SC joining the health network with no prior professional connections.
   \item  PC who prefer SC outside the network can undermine the effort to keep patients in the organization's health network.
\end{enumerate}

\subsection{Artificial Intelligence in Healthcare}
The healthcare industry has experienced a significant impact from the rapid advancement of Artificial Intelligence (AI) systems\cite{rajpurkar2022}. With the strong growth of multimodality data, data analytics, and AI play an increasingly important role in healthcare. Expert AI systems are employed to help physicians diagnose diseases based on symptoms\cite{aboueid2019}, while more advanced AI systems employ Electronic Health Records (EHRs) to predict medical events\cite{Jagannatha2016}.

AI is applied across a range of healthcare domains, including computer vision for medical imaging and natural language processing for EHR data analysis. Reinforcement learning is being explored in the context of robotic-assisted surgery, while generalized deep learning methods have been successfully applied to genomics research~\cite{ravi2016deep, esteva2019guide}. Moreover, knowledge bases linking diseases and symptoms extracted directly from EHRs can be leveraged to support clinical decisions and probabilistic models for self-diagnostic symptom checkers~\cite{rotmensch2017learning}. Additionally, complex networks and graph theory have been employed to study healthcare dynamics, such as the effects of the COVID-19 pandemic on acute patient care and to represent patient journeys in a network~\cite{kohler2022using}.

\subsection{Recommender Systems and Graph Neural Networks in Health Care}
The use of AI systems in the healthcare sector has been on the rise, particularly for supporting medical decisions such as clinical and medication predictions. EHRs are a valuable source of data for these tasks, but their heterogeneity and complexity pose challenges for AI algorithms. To address these challenges, state-of-the-art deep learning methodologies have become the go-to approach in recent years, as they are able to handle the complexity and diversity of EHRs data~\cite{wang2019mcpl,song2020local, kodialam2020deep}. In particular, deep neural network models have been developed to predict patients' future medical needs based on their clinical records extracted from EHRs. These models have been shown to be effective in recommending necessary diagnostic procedures for patients~\cite{noshad2020clinical}. Research in this area has also focused on recurrent neural networks (RNNs) and graph neural networks (GNNs), which are used to retrieve sequential and dynamic information from EHRs. Among these, GNNs are considered state-of-the-art and are used to extract predictive power from the connectivity in EHRs~\cite{li2020graph,li2020knowledge, liu2020heterogeneous}.

One area of research that has received much attention is diagnosis prediction, which involves predicting patients' future diagnoses based on their medical history. GNNs have been found to outperform RNNs in this task, as they can better handle the heterogeneous and ever-changing information in EHRs~\cite{li2020graph,li2020knowledge, liu2020heterogeneous}. Furthermore, a combination of GNNs and RNNs has been shown to increase performance in predicting patients' next prescriptions~\cite{liu2020hybrid}. Graph-based representations and graph embeddings have also been studied to make EHRs more easily interpretable in AI systems~\cite{choi2017gram, yue2020graph}. In addition to these developments, GNNs have been applied to COVID-19 studies. Moreover, GNNs have been applied to current health crises, such as improving the prediction of weekly COVID-19 cases~\cite{fritz2022combining} and inferring potential drugs for COVID-19 treatment~\cite{hsieh2021drug}.

%\section{Data Collection and Data wrangling}
\section{Materials and Methods}
%This section describes the datasets, data preparation, and methodology used to analyze large-scale patient consultation transaction data from a Portuguese private healthcare provider. Exploratory data analysis revealed expected trends in patient age distribution, the number of consultations throughout the year, and the number of appointments for each specialty. Further analysis showed a power law distribution in the number of physicians one patient has had appointments with. Joining the information about the physicians with the consultations information led to the construction of a weighted referral bipartite network and a professional network of physicians, which revealed potential inefficiency concerns for lack of involvement in the referral process.

\subsection{Data Collection and Data Preprocessing}

For the purpose of evaluating referral patterns and validating our hypotheses, we analyzed patient consultation transaction data from a Portuguese private healthcare provider. This dataset consisted of over 12 million consultations and roughly 1.4 million patients, with information on patient demographics (e.g., gender, age, nationality, and region), the consulting physician, consultation date, and hospital location. Additionally, we received data on each physician's demographic characteristics, including gender, age, education, hospital, and first (and second, if applicable) specialty. A total of 3,632 physicians were identified, including 389 primary care physicians (PCs) and 1,313 specialty care physicians (SCs). The remaining 1,930 physicians had unknown specialties and were associated with seven hospitals.

This study was based on the re-use of an anonymized research dataset provided by the health provider after the risk assessment and data sharing agreement was evaluated and approved. The dataset used was sufficiently aggregated and anonymized to prevent the identification of individual subjects. Furthermore, all methods were carried out in accordance with relevant guidelines and regulations, which allow for the documentation and processing of pseudonymized patient and service provider data for specific research purposes.

%All data collection procedures methods were conducted in accordance with European data protection standards and were kept confidential. The risk assessment of data usage and informed consent was obtained from the data provider, José de Mello Saúde. For this study, we only used the necessary data, namely physician database data (gender, age, and nationality) and patient demographic information (for descriptive analysis). All results were presented in such a way as to prevent the identification of any individual based on the data and information provided.

\subsection{Exploratory Data Analysis}

\begin{figure}[!htb]
\centering
\includegraphics[width=\linewidth]{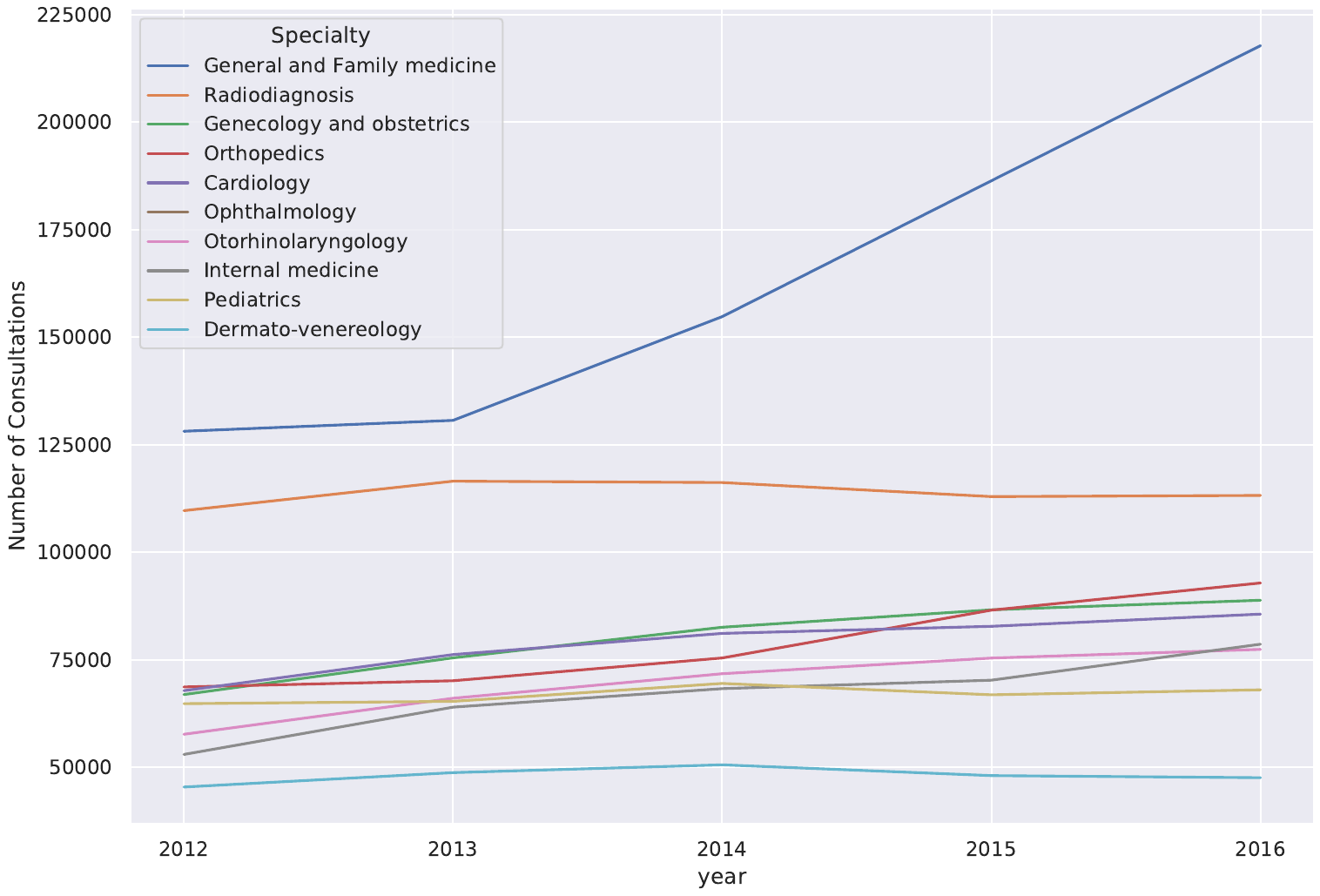}
\caption{\textbf{The number of appointments per specialization has varied over the years, with the nine most requested specialties.} In 2012, there was a marked increase in family physician consultations, while other specialties saw slower growth. It's also noteworthy that radiology has consistently had numerous consultations each year.}
\label{fig:especializacao-tempo}
\end{figure}

In our initial data exploration, we examined expected trends such as patient age distribution and the number of consultations across specialties over time. We found that the patient age distribution showed a higher percentage of younger (under 18) and older (over 65) patients, as expected due to the increased need for healthcare in those age groups.  Additionally, the number of consultations throughout the year was higher in the colder months, as seen in the Supporting Information Section~(\ref{suporting_info}). However, the number of appointments for each specialty followed different trends. Figure~\ref{fig:especializacao-tempo} illustrates the trend in consultations per specialty over time, with a notable increase in PC consultations in 2012. In particular, radiology consistently had a high number of consultations each year, likely due to its broad applicability across various medical needs.

%To gain a better understanding of the data, we investigated expected trends. For example, the patient age distribution showed a higher percentage of younger (under 18) and older (over 65) patients, as expected due to the increased need for healthcare in those age groups. Additionally, the number of consultations throughout the year was higher in the colder months, as seen in the Supporting Information Section~(\ref{suporting_info}).

%The number of appointments for each specialty followed different trends. Figure~\ref{fig:especializacao-tempo} shows a sharp increase in the number of family physician consultations in 2012, while the rise in other specialties was relatively slow. It is also noteworthy that the specialty of radiology had a high number of consultations per year.

\begin{figure}[!htb]
\centering
\includegraphics[width=1.2\linewidth]{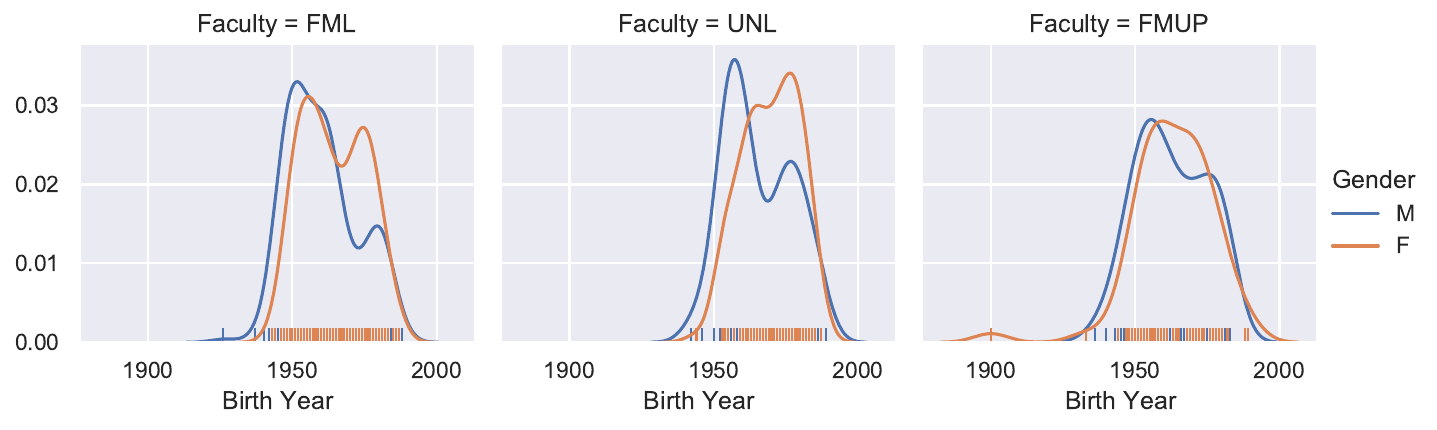}
\caption{\textbf{Figure 2. Age distribution of physicians from the three most frequented Faculty of Medicine:}  University of Lisbon (FML), University of Porto (FMUP), and Nova University of Lisbon (UNL). Both FML and UNL observed a significant increase in female medical students during the 1970s and 1980s.
}
\label{fig: gender-age-physicians}
\end{figure}

Figure~\ref{fig: gender-age-physicians} presents the age and gender distribution of physicians, demonstrating an increase in the percentage of women pursuing medicine in the 1970s and 1980s, likely as a result of socio-political changes in Portugal in 1974 at the end of a forty-year dictatorship. This trend may impact referral patterns, as previous research suggests that gender dynamics can influence professional interactions~\cite[Chapter~4]{highereducationportugal}.

%Analysis of the gender distribution of physicians based on their age reveals interesting trends. Figure~\ref{fig: gender-age-physicians} shows the birth year distribution of physicians for the top 3 most frequented medical schools, divided by gender. The data suggests that in the 1970s and 1980s, the percentage of women pursuing medicine increased. The phenomenon likely correlates with the end of a forty-year dictatorship in Portugal in 1974. After this event, women began to outnumber men in higher education institutions, not just in the health sciences but in many other academic fields~\cite[Chapter~4]{highereducationportugal}.

%Further analysis revealed that the number of physicians a patient had appointments with followed a power law distribution (see Figure~\ref{fig:power-law}). Many patients only consulted with one physician, so the data related to these patients was not relevant to exploring the referral mechanism.

\begin{figure}[!htb]
\centering
\includegraphics[width=\linewidth]{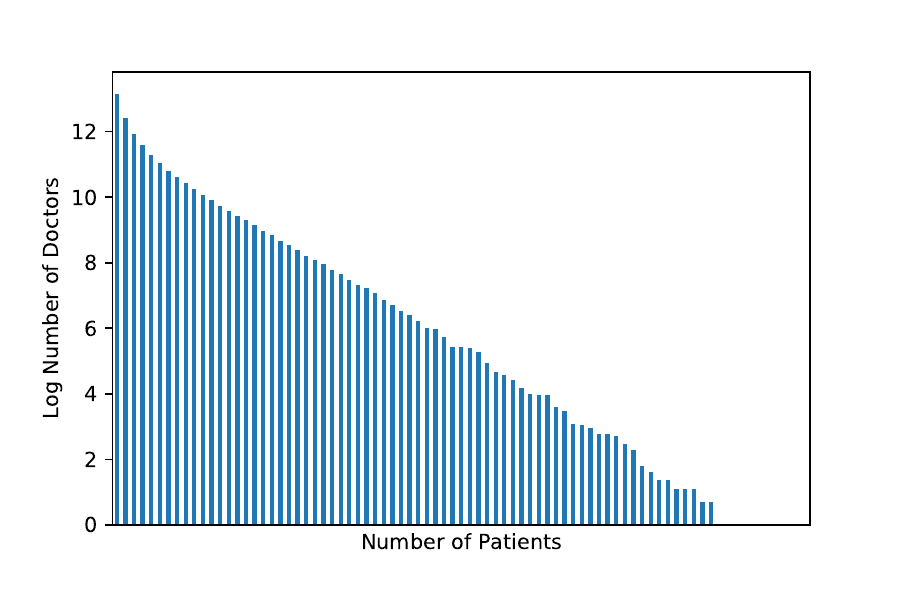}
\caption{\textbf{Number of physicians consulted by patients in log scale}, revealing a power-law pattern: a small segment of patients consult a large number of physicians, while the majority consult between one to three physicians
}
\label{fig:power-law}
\end{figure}
\begin{figure}[!htb]
\centering
\includegraphics[width=\linewidth]{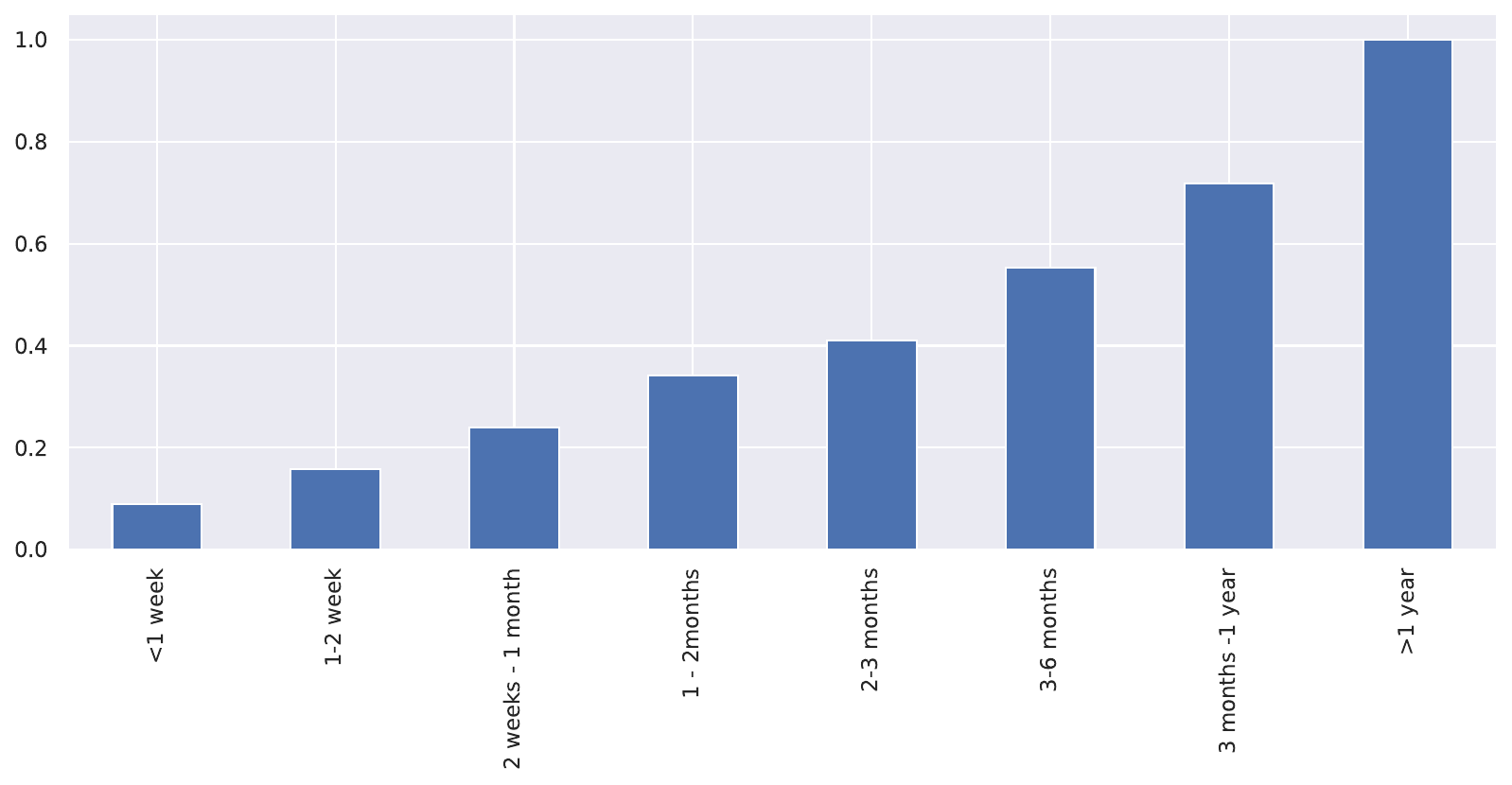}
\caption{\textbf{Cumulative distribution of the time interval of the physician-physician interactions over time}, based on consecutive patient appointments from primary to specialist care.  Approximately 22\% of interactions occurred within a month, providing the basis for the construction of the referral network.}
\label{fig: cumulativa}
\end{figure}

In further analysis, we found that the number of physicians a patient consulted with followed a power-law distribution, indicating a small group of physicians received a significant amount of referrals (see Figure~\ref{fig:power-law}). We defined a physician-physician interaction as a pair of consecutive appointments with the same patient, with the initial consultation being a PC and the following one an SC. Figure~\ref{fig: cumulativa} shows the cumulative percentage of these interactions over various time intervals, providing insights into the temporal dynamics of the referral process.

%Figure~\ref{fig: interecce-esp-perc} further explores this idea.

%To better understand physicians' referrals and social networks, we computed several summary network metrics. Our findings reveal that the degree distribution of nodes in the referral network follows a power-law distribution, suggesting that a small number of physicians receive a significant amount of referrals, while most physicians either refer to or receive referrals from a small number of colleagues.

%We analyze referrals by joining information about physicians and consultations. Figure~\ref{fig: cumulativa} shows the cumulative percentage of interactions over time intervals. An interaction is a pair of appointments with the same patient, the first being a primary care consultation and the second a specialty appointment. We only consider interactions with the minimum time interval between the two appointments (e.g., a primary care appointment in January followed by a cardiology appointment in February is an interaction with a one-month break). Figure~\ref{fig: interecce-esp-perc} further explores this idea.

\iffalse
\begin{figure}[!htb]
\centering
\includegraphics[width=0.85\linewidth]{Imagens/interesse_especializacao_percentagem.pdf}
\caption{\textbf{Percentage of interactions of each specialty that occurred per time interval}. Cardiology and Radio-diagnosis have the highest rate of appointments within a week of a primary care appointment---around 13\% of all interactions. Conversely, in Orthopedics, it is more likely to have an appointment with a specialist one or two months after the primary care consultation. It is an interesting research question whether the differences in average time between appointments of some specialties compared to others can be explained by perceived urgency.}
\label{fig: interecce-esp-perc}
\end{figure}
\fi

\subsection{Methodology}

%This section provides a detailed description of the methods used in the study, which involved the construction of two graphs, a referral network and a professional social network of physicians, and the application of Graph Neural Networks (GNN) to predict referrals based on social network information. The study used 1134 nodes and 27,743 edges to train the models and was tested with the same number of nodes and 30,825 edges. The study also considered gender, age, and centrality measures as key features.

%This section describes the construction of two graphs, a referral network and a social network of physicians, and the use of Graph Neural Networks (GNN) to predict referrals using the social network information. The models were trained with 1134 nodes and 27,743 edges, tested with the same number of nodes and 30,825 edges and used features such as gender, age, and centrality measures.

\subsubsection{Medical Referrals from a Network Perspective}
\label{sec: medical referrals as a social network}
\begin{figure}
     \centering
     \begin{subfigure}{0.49\textwidth}
         \centering
         \includegraphics[width=\textwidth]{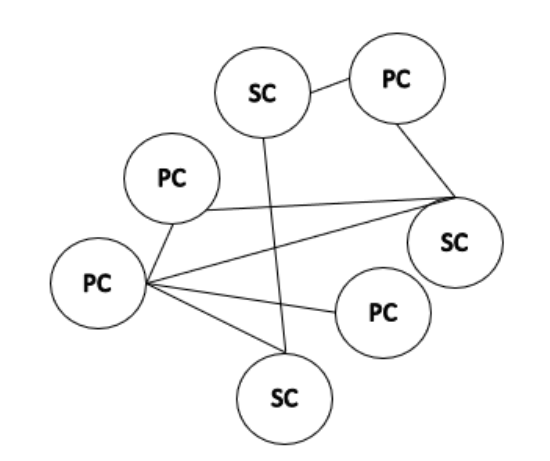}
         \caption{\textbf{Depiction of the two constructed networks primary care physicians (PC) to specialists (SC):}
          (a) The social network models interactions between physicians, connecting those with professional ties. (b) The referral network reflects referrals from PC to SC.}
         \label{fig:social}
     \end{subfigure}
     \hfill
     \begin{subfigure}{0.49\textwidth}
         \centering
         \includegraphics[width=\textwidth]{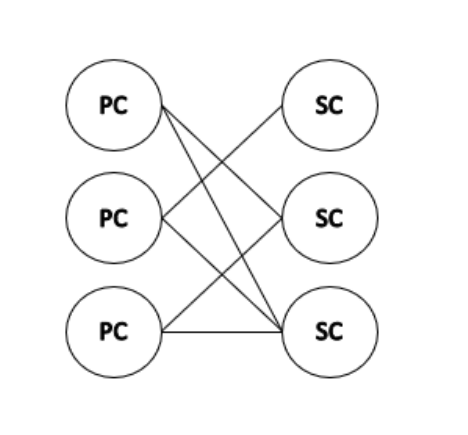}
         \caption{Referral: A primary care (PC) physician can refer a patient to a specialty physician (SC) if the patient had an appointment with the PC up to one month before an appointment with the SC.}
         \label{fig:referral}
     \end{subfigure}
        \caption{\textbf{Representation of the two graphs constructed}. The social network in (a) models the interactions between physicians. Two physicians are connected if they have a professional or educational relationship. Similarly, in (b), the referral network models the referrals made by primary care physicians (PC) to specialists (SC).}
        \label{fig:graphs}
\end{figure}

Following data preprocessing and initial analysis, we constructed a weighted bipartite referral network and a professional social network of the physicians (see Figure \ref{fig:graphs}). More specifically, consistent with the existing literature, the referral network was built using  consecutive patient consultations to a PC and a SC occurred within a month ~\cite{An2018,Landon2018}. Each link in the referral network connects one PC to another SC, with 294 PC physicians and 839 SC physicians connected via 34,249 edges. The edge weight on the referral network represents the number of patients that the PC physician refers to the SC physician. However, not all physicians participated in the referral network, suggesting potential inefficiencies in the system.

%With the data preprocessing and initial analysis complete, we constructed a \emph{referral weighted bipartite network} and a \emph{professional network} of the physicians (see Figure\ref{fig:graphs}). Following common standards used in Landon et al.~\cite{Landon2018} and An et al.~\cite{An2018}, the referral graph was built using only the interactions that occurred within a month. The links represent referrals from Primary Care to Specialist Care physicians, so links only connect one PC physician to another SC physician. In the referral graph, 294 PC physicians are connected with 839 SC physicians through 34,249 edges. The edge weight on the referral network indicates the number of patients that the PC physician refers to the SC physician. 
To complement the referral network, we also use information about the physicians' educational and professional backgrounds, including where they studied, did their residency, and currently work, to construct a professional network of the physicians based on their shared backgrounds.

%The social network was created using information about the physicians, such as where they studied, where they did their residency, and where they currently work. These connections are based on shared educational and professional backgrounds.

\subsubsection{Referral Prediction using Graph Neural Networks (GNN)}
Node embeddings derived from graph-structured data provide low-dimensional vector representations for each node, based on its graph neighborhood \cite{Grover2016}. This approach is useful for machine learning applications such as node classification, clustering, and link prediction. To generate node embeddings for the referral network, we adopted the GraphSAGE model\cite{hamilton2017inductive}. The GraphSAGE model is an unsupervised representation learning model that uses the graph structure and node features to learn node embeddings from large graphs. It can leverage node attributes to learn the structure of each node's neighborhood jointly with the distribution of node features in the neighborhood~\cite{hamilton2017inductive}. The GraphSAGE model is useful in solving classification tasks, such as predicting whether a node pair will likely co-occur in a random walk on the graph. Positive node pairs are generated from performing random walks, and negative node pairs are generated randomly according to a distribution. With this simple classifier, the model can learn inductive mappings from attributes of nodes and their neighbors to node embeddings while preserving the structural similarities of nodes and their features. These GraphSAGE node embeddings serve as the base for the link prediction tasks we performed on the referral network to test the hypothesis that the social features of physicians influence the referral process. We used the Stellargraph Library, a Python library for machine learning on graphs and networks that had already implemented the GraphSAGE model for link prediction in node embeddings~\cite{StellarGraph}.

From the professional social network of physicians, we computed three centrality measures for each node: betweenness, eigenvector, and degree centrality, and added them as features to GraphSAGE  to accomplish the link prediction task based on the referral network. Firstly, we learned an unsupervised graph representation of the referral network.  We trained the GNN with 1,134 nodes and 27,743 edges and tested it with the same number of nodes and 30,825 edges. Features used were physician gender and age, plus the centrality features for the social network-aware experiment. We trained for 20 epochs, layer sizes of 20 by 20, and a dropout rate of 0.3. A neural network model was trained for link prediction, optimizing with Adam (learning rate of 1e-3) over the binary cross-entropy loss. In comparison, we also evaluated the link prediction task on the referral network in the \emph{absence} of social network information.

To ensure that the results were not dependent on the node embedding model, we also performed similar tasks by training two baseline models, namely Attri2Vec\cite{Attri2Vec} and Node2Vec\cite{Node2Vec}. In contrast to GraphSAGE, the Attri2Vec node embeddings are learned by performing linear/non-linear mapping on node content features. The model generates node pairs of node target and node context and uses them to learn the representation of the target node with the existence of the context node. This is done using a deep learning model that minimizes the binary cross-entropy loss function about the predicted pair and true pair labels using stochastic gradient descent.

Meanwhile, we used the Node2Vec embedding model to generate node embeddings and performed the link prediction task using Logistic Regression. Node2Vec is a graph embedding algorithm that produces a vector representation of a node based on random walks in the graph. The algorithm samples neighborhood nodes through deep random walks, performing a biased random walk procedure to efficiently explore diverse neighborhoods, similar to Word2Vec. The Node2Vec model has four key parameters: the number of walks, the walk length, the return hyperparameter $P$, and the in-out hyperparameter $Q$. We generated 10 random walks from each node in the graph, with each walk containing 80 nodes. The return and in-out hyperparameters were set to 1. The output dimension of the embeddings was set to 128. Table~\ref{tab:models-explained} provides a summary of the models used in our study and their respective setups.

\begin{table}
\centering

    \begin{tabular}{|c|c|c|c|}
        \hline
         \textbf{Model} & \textbf{Train set numbers} & \textbf{Test set numbers} & \textbf{Hyperparameters}  \\
         \hline
        GraphSAGE & \Centerstack[l]{1134 nodes \\
   27743 edges} & \Centerstack[l]{1134 nodes \\
   30825 edges}  &   \Centerstack[l]{20 by 20 layer network \\
   Epochs 20\\ Dropout 0.3 \\ Adam learning rate 1e-3}  \\
         \hline
         Attri2Vec  & \Centerstack[l]{1134 nodes \\
   27743 edges} & \Centerstack[l]{1134 nodes \\
   30825 edges}  &   \Centerstack[l]{128 layer network \\
   Epochs 10\\ No regularization \\ Adam learning rate 1e-2}  \\
         \hline
         Node2Vec & \Centerstack[l]{1134 nodes \\
   4628 edges} & \Centerstack[l]{1134 nodes \\
   6848 edges}  &   \Centerstack[l]{Number of walks 10 \\
   Wlak length 80\\ $P = Q = 1$ \\ Adam learning rate 1e-2}  \\
         \hline
    \end{tabular}
    \caption{Summary of all the models trained. Node2Vec model served as a baseline.}
    \label{tab:models-explained}
\end{table}

\subsubsection{Node embeddings and Feature importance}

The previous section presented the GraphSAGE and Attri2Vec models, which produced node embeddings relevant to identifying patterns in the referral graph and demonstrating the importance of physicians' social features in the referral mechanism. In this section, we visualization the node embeddings by reducing their dimensionality from 20 to 2 using U-MAP~\cite{mcinnes2018umap}, T-SNE~\cite{van2008visualizing}, and ISOMAP~\cite{tenenbaum2000global} techniques. Our aim is to identify similar structures and explanations for the medical referrals' behavior in the node embeddings of the referral graph produced by both models.

To further ensure the robustness of our findings and the significance of social features for referral prediction, we use the Shapley Additive exPlanations (SHAP) feature importance framework that predicts the existence of links in the referral graph to evaluate the importance of social features for referral prediction in a model-agnostic-wise manner~\cite{NIPS2017_7062}. The SHAP framework is a cooperative game theoretic approach known as the Shapley value that assigns each feature an importance value for a particular prediction by identifying a new class of additive feature importance measures~\cite{shapley1953value}. Despite being a state-of-the-art explainer model with desirable properties, it makes a few critical assumptions that we need to consider when applied to our data. For instance, it assumes that observations are independent, which is not the case across time, physicians, and patients. Additionally, the additive structure of feature interaction has little support from real data. Therefore, SHAP analyses must be cautiously examined as local approximations~\cite{slack2020fooling}.

More specifically, we trained a binary neural network classifier that uses the existence or non-existence of a link in the referral graph as the positive and negative labels, respectively, together with the five features (age, gender, and three centrality measures) as the input. The classifier's architecture consists of two hidden layers, with five nodes each, a fully connected network with ReLu activations, and a sigmoid activation in the last layer. The network model is optimized with Adam (learning rate of 1e-3) over the binary cross-entropy loss for 30 epochs. With this trained model, we assess feature importance through the SHAP framework.

%The model trained was a simple classifier with the label being the existence or non-existence of a link in the referral graph and the features being the five features included in the GraphSAGE Model (age, gender, and the three social features) from the two physicians whose association was being classified. We use 6164 balanced supervised examples to build the model and use 30 \% of the examples for test purposes. The model's architecture was a two hidden-layer, with five nodes each, a fully connected network with ReLu activations, and a sigmoid activation in the last layer. The network model was optimized with Adam (learning rate of 1e-3) over the binary cross-entropy loss for 30 epochs. The accuracy of the model was near 68 \%, and the loss was about 0.62. With the trained model, we assess feature importance through the SHAP framework.

To get a better sense of the impacts of age and gender in the link prediction results, we built another model --- with the same architecture as the previous neural network classifier but with engineered features for age and gender that replaced the other four features. The new model has all the social metrics for both $g$ and SC physicians as features and has two more new features $\Delta a$ and $g$ for the age difference between physicians and gender combinations computed as
\begin{equation*}
    \Delta a = |{\text{age}_{PC} - \text{age}_{SC}}|
\end{equation*}
where $\text{age}_{PC}$ is the estimated age of the primary care physician referring the patient and $\text{age}_{SC}$ is the estimated age of the specialist physician being referred to, and $g$ is a categorical variable indicating the possible gender combinations in the referral relationship as
\begin{align*}
    g(PC,SC)=
    \begin{cases}
      0 & \text{if  PC is a female and SC is a female}\\
      1 & \text{if  PC is a male and SC is a male}\\
      2 & \text{if  PC is a male and SC is a female}\\
      3 & \text{if  PC is a female and SC is a male}.
    \end{cases} 
\end{align*}

\section{Results}

%This study found that including information about the social network of physicians improves the prediction of referrals for all models, with the Attri2Vec model producing the best results. Visualizations of the node embeddings show that with professional network features, the node embeddings present more structure, indicating that the professional network features can bear information about referrals. Additionally, the gender of the physicians is a relevant feature for predicting referrals, with an underlying pattern of physicians referring to physicians of the same gender.

To summarize, this section examined the impact of including information about the professional network of physicians on the referral predictions. We employed various graph-based link prediction algorithms and found that incorporating such data improves the predictive power of all models. he Attri2Vec model, in particular, demonstrated most favorable results. Further analysis of the node embeddings revealed that incorporating professional network features resulted in more structured embeddings, indicating that these features hold relevant information about referrals. Additionally, we found that the gender of physicians played a significant role in predicting referrals, with physicians displaying a tendency to refer to colleagues of the same gender.

%\subsection{Physicians cluster together and mainly refer to a small number of colleagues}

%Referral and professional networks of physicians
%We computed summary network metrics for the resulting referral and social networks. For example, the degree of nodes in the referral network follows a power-law distribution. This means that: (1) few physicians receive a massive amount of referrals (if the node is an SC) or that they refer to many specialty physicians (if the node is a PC), (2) the vast majority of SCs only receive referrals from few PCs or PCs refer to a few SCs. Such observations are subject to several possible explanations. The SC with high in-degree can be those with high popularity, while PC with low out-degree may have relatively limited social contacts. Meanwhile, we obtained the average clustering coefficient for the referral network (0.149) to measure the fraction of the number of observed triangles to the total number of possible triangles in the network. This represents an essential precondition for the referral network to exhibit a small-world structure and suggests that physicians in the referral network have a higher tendency to cluster together. 

\subsection{Including information about the professional social network of physicians improves the prediction of the referrals for all models}
Table~\ref{tab:link-results} presents the referral link prediction results obtained from the out of sample test set. Our results demonstrate that including information about physicians' professional social networks improves the accuracy and loss function of all GNN models. The Attri2Vec model shows the greatest improvement, confirming the contribution of these features to the referral prediction. Our analysis of node embeddings further supported these findings, as incorporating professional network features resulted in more structured embeddings, indicating that these features provide valuable information about referrals. As such, the Attri2Vec model was selected for further analysis in the subsequent section.

%The results of the medical referral link prediction for all the models are presented in  Table~\ref{tab:link-results}. We show the accuracy and loss on the test set for all the model setups considered. Including information about the social network of physicians improves the prediction of the referrals, regardless of the graph-based link prediction algorithm used. Yet, it is the Attri2Vec Model that produced the best results in terms of link prediction accuracy.
%These findings suggest that the information on physicians' social relationships improves the model's predictive power. Using the GraphSAGE model, accuracy increases by about 0.18 with the added information about physicians' social networks; as expected, the loss decreases as well. Similarly, the results concerning the Attri2Vec model show  an increase in the accuracy and a decrease in the loss function in the presence of social features, confirming the contribution of these features to the referral prediction. As the Attri2Vec performs better on the network with social information, the node embedding analysis presented in the next section will focus on the embeddings of this model.

\begin{table}[!htb]\renewcommand{\arraystretch}{2}\addtolength{\tabcolsep}{-1pt}
\centering

    \begin{tabular}{c|c|clc}
        \hline
         \textbf{Model} & \textbf{Professional Network} & \textbf{Out-of-sample loss} & \textbf{Out-of-sample accuracy}  \\
         \hline
         \hline
         \multirow{2}{*}{GraphSAGE} & Without information & 1.0629  &   0.5232 \\
         & With information & 0.6939 & 0.7072\\
         \hline
         \multirow{2}{*}{Attri2Vec} & Without information & 0.6790 &   0.6876 \\
         & With information & \textbf{0.6254} & \textbf{0.7268}\\
         \hline
         \multirow{2}{*}{Node2Vec} & Without information & --- &   0.5828 \\
         & With information & --- & 0.6396\\
         \hline
    \end{tabular}
    \caption{Link prediction task metrics, when social network features are added to referral information with different model embeddings --- GraphSAGE, Attri2Vec, and Node2Vec all improve with professional network information.}
    \label{tab:link-results}
\end{table}

\subsection{Node embeddings show improved structure with the inclusion of professional network features}

Although the dimensionality reduction techniques UMAP, t-SNE, and ISOMAP employed for Attri2Vec node embeddings are fundamentally distinct, they exhibit some shared structural characteristics and observations. In this paper, we primarily concentrate on the UMAP results, while providing the outcomes of the other dimensionality reduction algorithms in the supplementary material.

\begin{figure}
\centering
\includegraphics[scale=0.5]{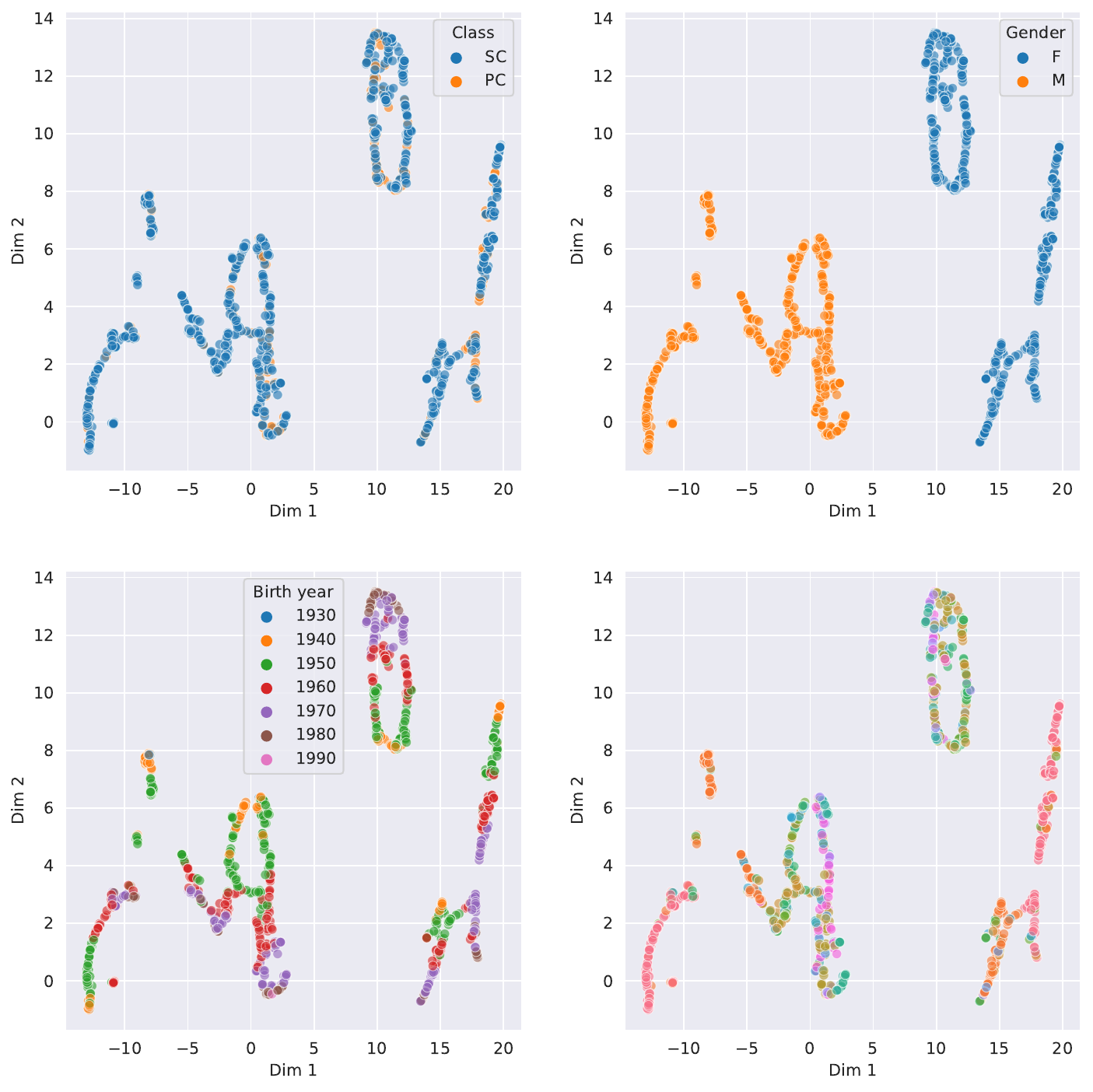}
\caption{\textbf{Two-dimensional visualization of node embeddings obtained from the Attri2Vec model using U-MAP,  incorporating social features}.Two-dimensional visualization of node embeddings from the Attri2Vec model using U-MAP, incorporating social features. Plots distinguish between PC and SC, gender, physicians' birth year, and hospital, with each cluster representing a group of physicians who refer within the group.}
\label{fig:umap-with}
\end{figure}
\begin{figure}
\centering
\includegraphics[scale=0.5]{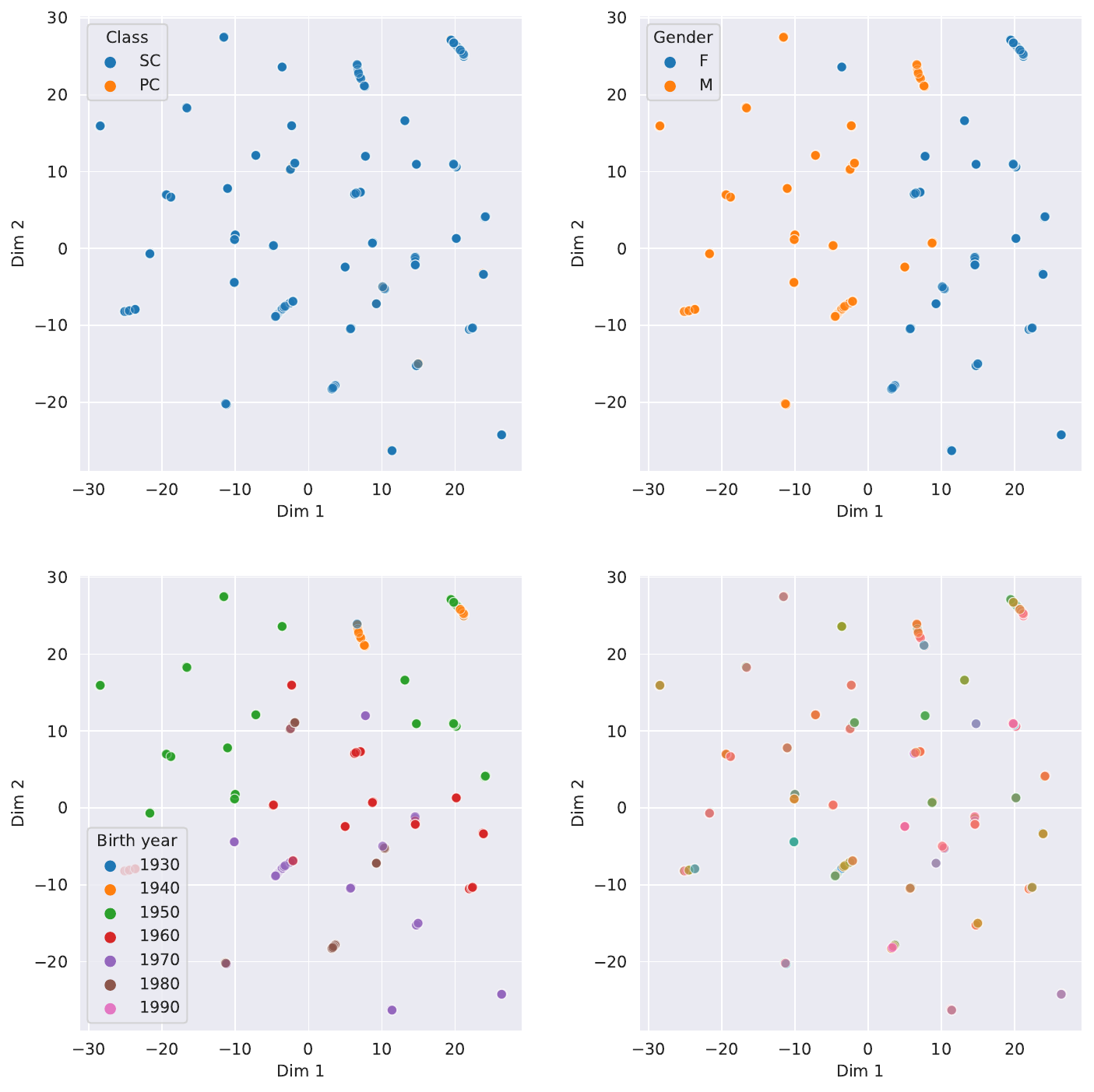}
\caption{\textbf{wo-dimensional visualization of node embeddings from the Attri2Vec model using U-MAP, without social features. The plots differentiate between PC and SC, gender, physicians' birth year, and hospital. These embeddings exhibit less structure than those incorporating social features, emphasizing their importance in referral prediction.}. The plots differentiate between PC and SC, gender, physicians' birth year, and hospital. These embeddings exhibit less structure than those incorporating social features, emphasizing their importance in referral prediction.}
\label{fig:umap-without}
\end{figure}

To investigate the impact of professional network features on the embedded representation of physicians within the referral network, we trained Attri2Vec (the optimal Graph Neural Network from Table~\ref{tab:link-results}) both with and without professional network features. The physicians' embeddings were then reduced to two dimensions, and the two-dimensional representations were visualized with (Figure~\ref{fig:umap-with}) and without professional network features (Figure~\ref{fig:umap-without}). Each physician's data point was colored according to distinct attributes of interest. We observed that the node embeddings exhibit greater structure when professional network features are incorporated (Figure~\ref{fig:umap-with}), elucidating the success of the link prediction task with the additional information. Conversely, the representations without the professional network information demonstrate an absence of structure (Figure~\ref{fig:umap-without}).

We validated the anticipated behavior that (1) referrals involve relationships between PC and SC, with clusters containing both types (Figures~\ref{fig:umap-with} and~\ref{fig:umap-without}, top-left), and (2) physicians working in the same hospital tend to cluster together when provided with professional network features, even if the model is uninformed of the hospital, each professional is affiliated with (Figures~\ref{fig:umap-with} and~\ref{fig:umap-without}, bottom-right). Therefore, we can conclude that professional network features can encapsulate this information.

Our findings demonstrate that professional network features significantly impact physicians' embedded representation within the referral network. Furthermore, we observed that professional network features can capture crucial information, such as the relationships between PC and SC and the clustering of physicians in the same hospital, even when the model lacks direct access to this information. These results underscore the potential of professional network features to enhance the accuracy of referral management for healthcare providers.

The perspectives of age and gender reveal interesting patterns. In both embeddings, there is a clear distinction between female and male physicians, with this structure being more pronounced in the model without professional network features. Two primary conclusions can be drawn: (1) The model incorporating social features utilizes information beyond age and gender, resulting in more complex and accurate representations of reality. (2) gender remains a relevant feature for referrals, indicating an underlying pattern of physicians referring to those of the same gender.

In terms of age, this principle is not as strongly applicable. In the embeddings from the model with social features, age does not appear to be a strong discriminator, while it is more distinct in the simpler model. These results emphasize the importance of professional network features for the referral mechanism and indicate that, without this social information, the embeddings are not robust enough to explain the referrals.

%The node embeddings generated using the ISOMAP and t-SNE techniques exhibit similarities in analyzing the four aforementioned dimensions. The outcomes can be found in the supplementary information section.

\subsection{Feature Importance and SHAP Values}

The SHAP values of the features employed in the simple neural network (NN) link prediction model reveal that the specialist physician's degree centrality is the most critical factor for predicting a referral. This is followed by the eigenvector centrality of the primary care physician and the eigenvector centrality of the specialist physician. The primary care physician's age and gender have negligible importance on referrals, but the age difference's significance is more pronounced when inferred features are utilized.

\begin{figure}
\centering
\includegraphics[scale=0.7]{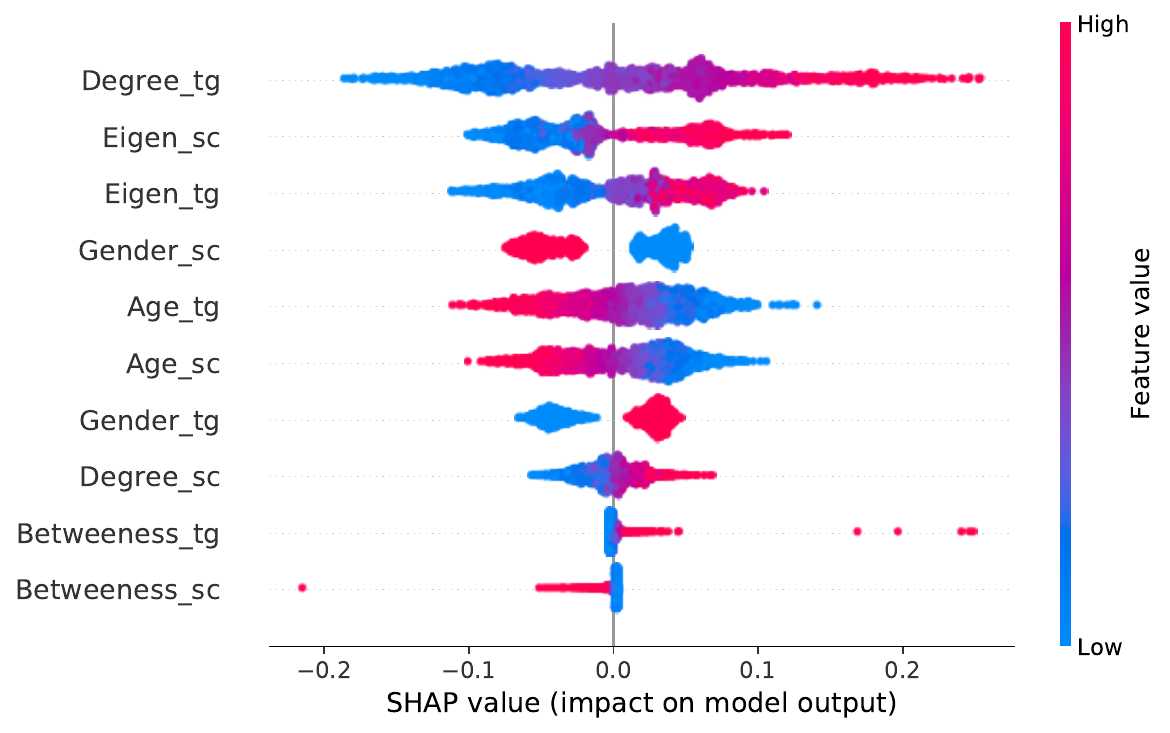}
\caption{\textbf{SHAP Values --- feature importance of a simple link prediction model using the network features and the age and gender of the physicians}. Suffixes \_sc (source) and \_tg (target) denote the type of physician to whom the features apply. The source physician refers to the PC, while the target physician refers to the SC.}
\label{fig:shap-normal}
\end{figure}

\begin{figure}
\centering
\includegraphics[scale=0.7]{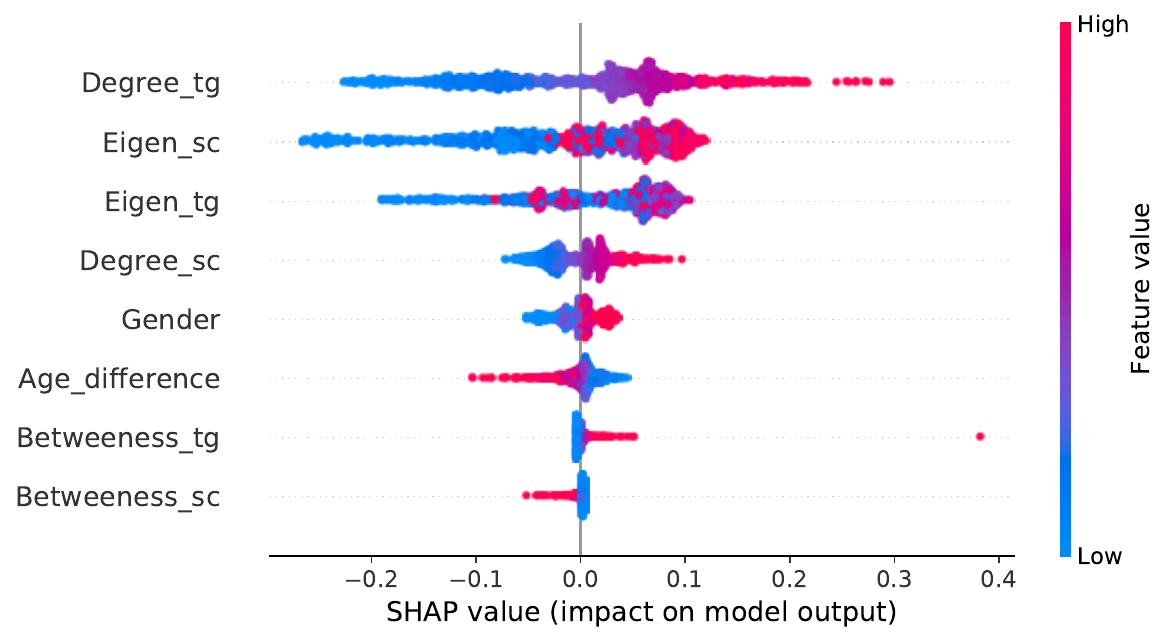}
\caption{\textbf{SHAP Values --- feature importance of a simple link prediction model, incorporating engineered features for age and gender of the physicians}. Suffixes \_sc (source) and \_tg (target) denote the type of physician to whom the features apply. The source physician refers to the PC, while the target physician refers to the SC.}
\label{fig:shap-new}
\end{figure}

Figures~\ref{fig:shap-normal} and~\ref{fig:shap-new} display the SHAP values of the features utilized in the simple NN link pre diction model. The specialist physician's degree centrality (\textit{Degree\_tg}) emerges as the most vital feature for predicting a referral, followed by the primary care physician's eigenvector centrality (\textit{Eigan\_sc}) and the specialist physician's eigenvector centrality (\textit{Eigan\_tg}).

These findings underscore the evident significance of the SC's degree centrality for referral predictions and the eigenvector centralities of both PC and SC. The SC's degree centrality plays the most prominent role in the referral process, while the PC's degree centrality is ranked among the least important features by SHAP. The simple count of occurrences denoted by degree centrality appears linear in terms of decision impact; however, the importance is substantially uneven for source and target nodes. In contrast, eigenvector centralities hold consecutive positions in the importance ranking but do not exhibit a linear influence on the model outcome.

The PC's age and gender exhibit diminishing importance on referrals. However, when engineered features (\textit{Age\_difference} and \textit{gender}) are incorporated into the model, the age difference's importance becomes more pronounced, and there is a tendency for physicians to refer to colleagues of the same age. The SC's number of connections within their professional network is the most critical factor in referral considerations. Additionally, the eigenvector centrality of a PC's relationships plays a significant role. Following these three factors, the PC's age and gender become less critical in determining referrals. In the model utilizing inferred features of age difference and gender, the age difference's importance is more accentuated, and there is a tendency for physicians to refer to those of the same age.

In summary, the centrality measures of specialist and primary care physicians within their professional network, as well as their mutual eigenvector centrality, play significant roles in predicting referrals. However, when inferred features are introduced into the model, age and gender become less crucial, while the age difference between physicians becomes more meaningful. This indicates a tendency for physicians to refer patients to colleagues around the same age. These findings underscore the importance of considering the professional network structure and specific attributes of physicians when attempting to predict referral patterns within the healthcare system.

\section*{Conclusions}

This study aimed to uncover hidden patterns in the referral system of physicians and to investigate the hypothesis that features derived from the professional network of physicians are essential for the referral of patients from PC to SC physicians. We provided evidence from a real dataset of a private European healthcare provider to support this intuitive statement. The node embeddings and link prediction tasks performed on the referral graph consistently indicated the importance of social information in physicians' referrals, regardless of the algorithm used. Physicians with more professional connections tended to receive more referrals, and those with significant connections had a higher likelihood of being referred. These findings were consistent across all models, with the Attri2Vec model showing the most promising results.

The findings of our study carry significant implications for the healthcare sector, particularly concerning patient referrals. The fact that a physician's professional network can influence referral patterns to such a degree suggests that there is more at play than purely clinical considerations in the referral process. This insight could lead to a reevaluation of current referral practices to ensure that they are patient-centric and not unduly influenced by social or professional networks. Moreover, the ability to predict referral patterns based on network features has potential applications in healthcare management. For instance, it could help in resource planning, identifying potential bottlenecks in patient flow, or even predicting the demand for certain specialties. This could ultimately lead to improved healthcare delivery and patient outcomes.

Future research should focus on developing an unbiased recommendation system that prioritizes patients' best interests in order to promote patient-centered care. However, it's important to note that our historical data, being biased, cannot be used directly for this purpose. Thus, additional data sources and methodologies may need to be explored to address this challenge.

\subsection*{Data Availability Statement}
Due to the sensitive nature of the data used in this study, which includes electronic health records of patients, it is not publicly available in order to ensure the privacy and confidentiality of the individuals involved. However, anonymized data or subsets of the data may be available from the corresponding author upon reasonable request and with permission of the relevant ethics committee. Please note that any data requests would be subjected to data use agreements and would be in compliance with the relevant data protection and privacy regulations.
\newpage
\iffalse
\input{04_Supporting_Information}
\newpage
\fi
%\nolinenumbers

\section*{Conflict of interest}
The authors declare that they have no known competing financial interests or personal relationships that could have appeared to influence the work reported in this paper.

\section*{Acknowledgments}
This work was partially supported by Fundação para a Ciência e a Tecnologia (UIDB/00124/2020, UIDP/00124/2020 and Social Sciences DataLab - PINFRA/22209/2016), POR Lisboa and POR Norte (Social Sciences DataLab, PINFRA/22209/2016), and NOVA LINCS (grant UIDB/04516/2020) and Carnegie Mellon Portugal
Program (CMU/TIC/0016/2021).
The authors would like to thank CUF Health for sharing the data and Oracle Cloud credits and related resources provided by Oracle for Research.

\bibliography{bibliography}

\iffalse
\section*{Data sharing statement}
The data set used in this study was provided by CUF Health, and it is prohibited from public release as it would violate patient privacy. We, therefore, are not permitted to release this data set.

\section*{Author contributions statement}

R.D. implemented the models and wrote the first draft, R.D., Q.H., C.S. conceived the experiments, R.D., Q.H., C.S. conducted the experiments, R.D., Q.H., C.S. analyzed the results.  All authors edited and reviewed the manuscript. 
\fi

\section*{Supporting information}
\label{suporting_info}

\begin{figure}[h]
    \centering
    \includegraphics[scale=0.5]{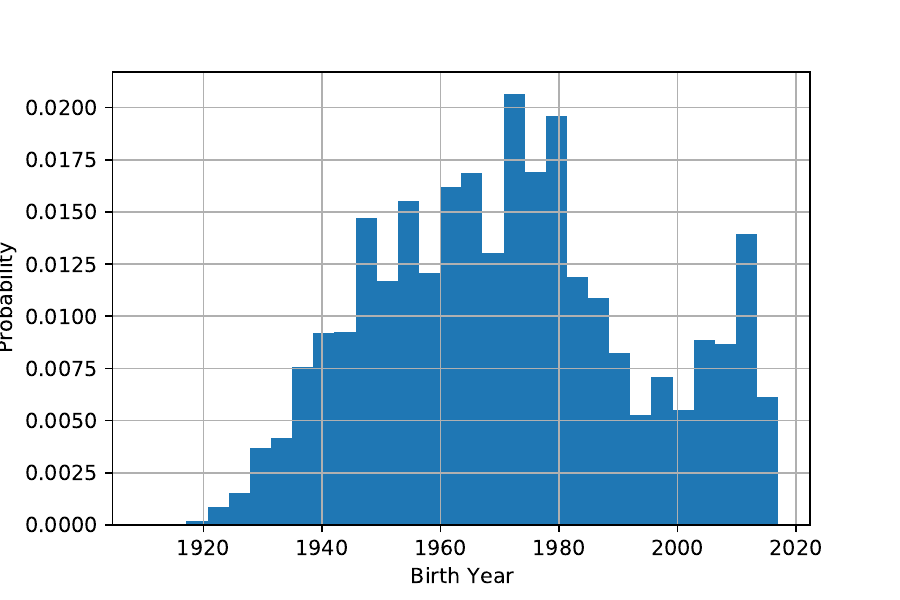}
    \caption{Distribution of birth year of the patients. Elder patients and the children are more common patients.}
 \label{idades_pat}
\end{figure}

\begin{figure}[h]
    \centering
    \includegraphics[scale=0.5]{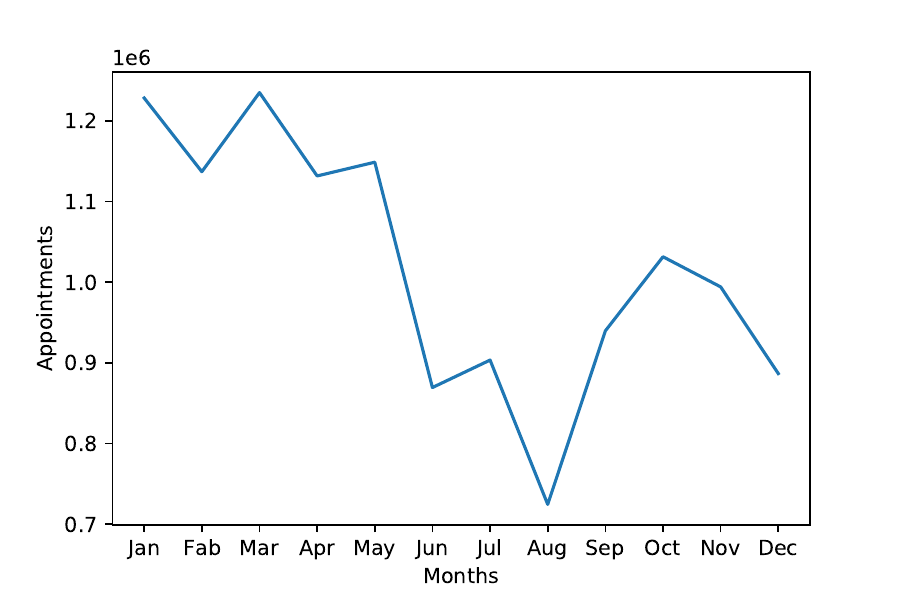}
    \caption{Number of consultations throughout the year. There are much more appointments in the cold months, and in December the number is also relatively low.}
 \label{idades_uni}
\end{figure}

\begin{figure}
    \centering
    \includegraphics[scale=0.5]{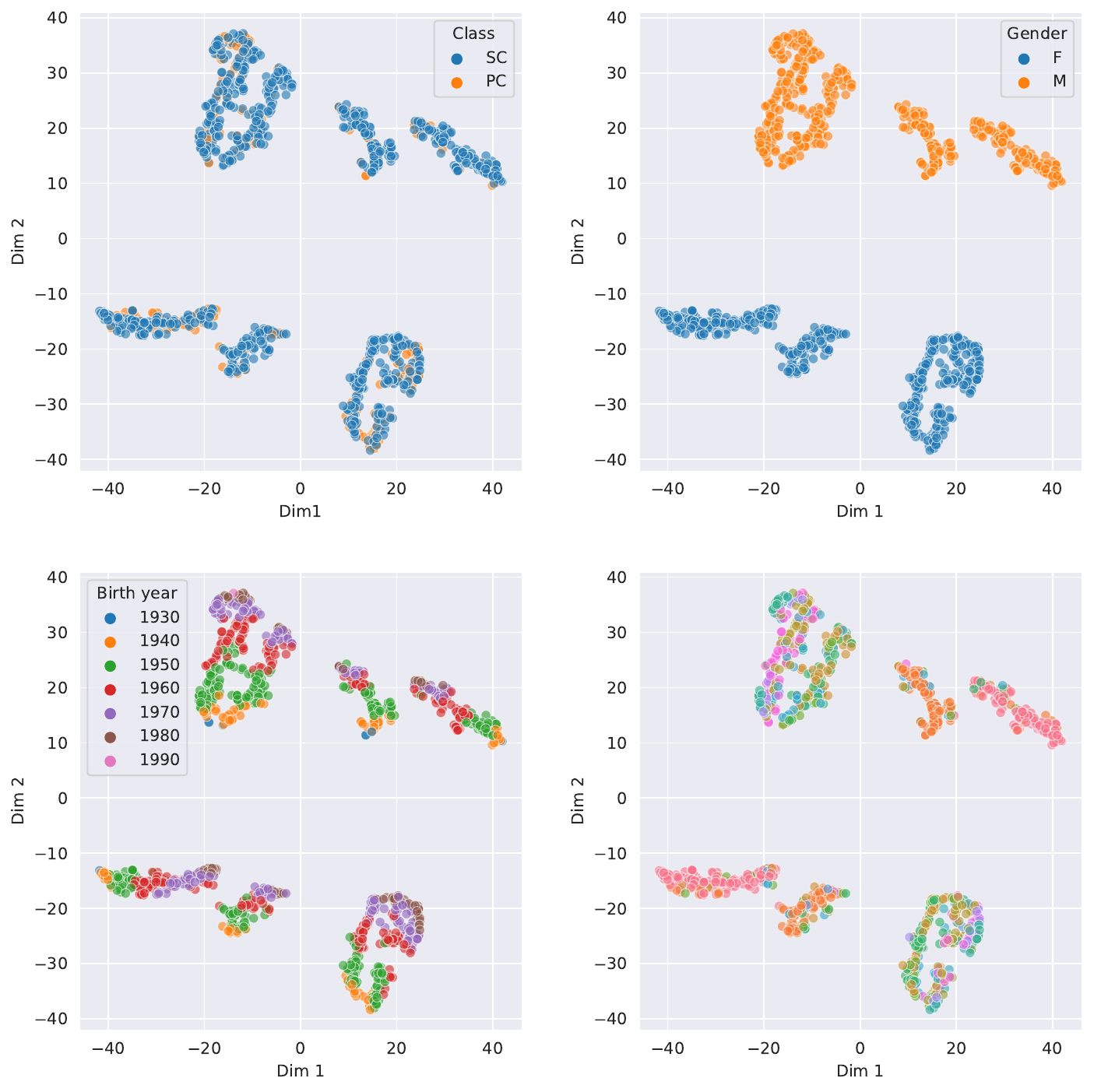}
    \caption{Node embeddings extracted from the Attri2Vec model with the social features reduced to two dimension using T-SNE with perplexity = 40. In the top left plot the node embedding separates primary care doctors (PC) from specialty doctors (SC); in the top right the embeddings are colored by gender. The embeddings naturally show some major clusters of doctors, which clearly are well separated by gender. As expected, the distribution of primary care doctors vs specialist doctors is similar throughout all clusters. The contrast of this embedding with the embedding of figure \ref{fig:tsne without_1} highlights the increase on the embeddings structures produced by the addition of the social features. In the bottom left, the node embeddings are colored by birth year of the doctors, whereas in the bottom right, the colors represent each hospital.The doctors who work in the same hospital are mostly represented in the same cluster--- this evidences, once more, the importance of the social features in representing this information in the model and in predicting the referrals. The birth year is also stratified by cluster, but is not sufficient to increase the performance of the referral predictor}
 \label{fig:tsne-with}
\end{figure}

%\paragraph*{S3 Fig.}
%\label{tsne without_1}
%{\bf  Node embeddings extracted from the Attri2Vec model without the social features reduced to two dimension using T-SNE with perplexity = 40} In the top left the node embedding separates primary care doctors (PC) from specialty doctors (SC); in the top right the embeddings are colored by gender. The gender is the most important feature in the decision on the model without the social features --- the model haven't the power to decompose the embeddings further.  In the bottom left the node embeddings are colored by birth year of the doctors whereas in the right, the colors represent each hospital.The pattern present in the ISOMAP visualizations is also present here. Without the social features, the information about hospital/ region where the clinician work not contribute to the prediction.

\begin{figure}
    \centering
    \includegraphics[scale=0.5]{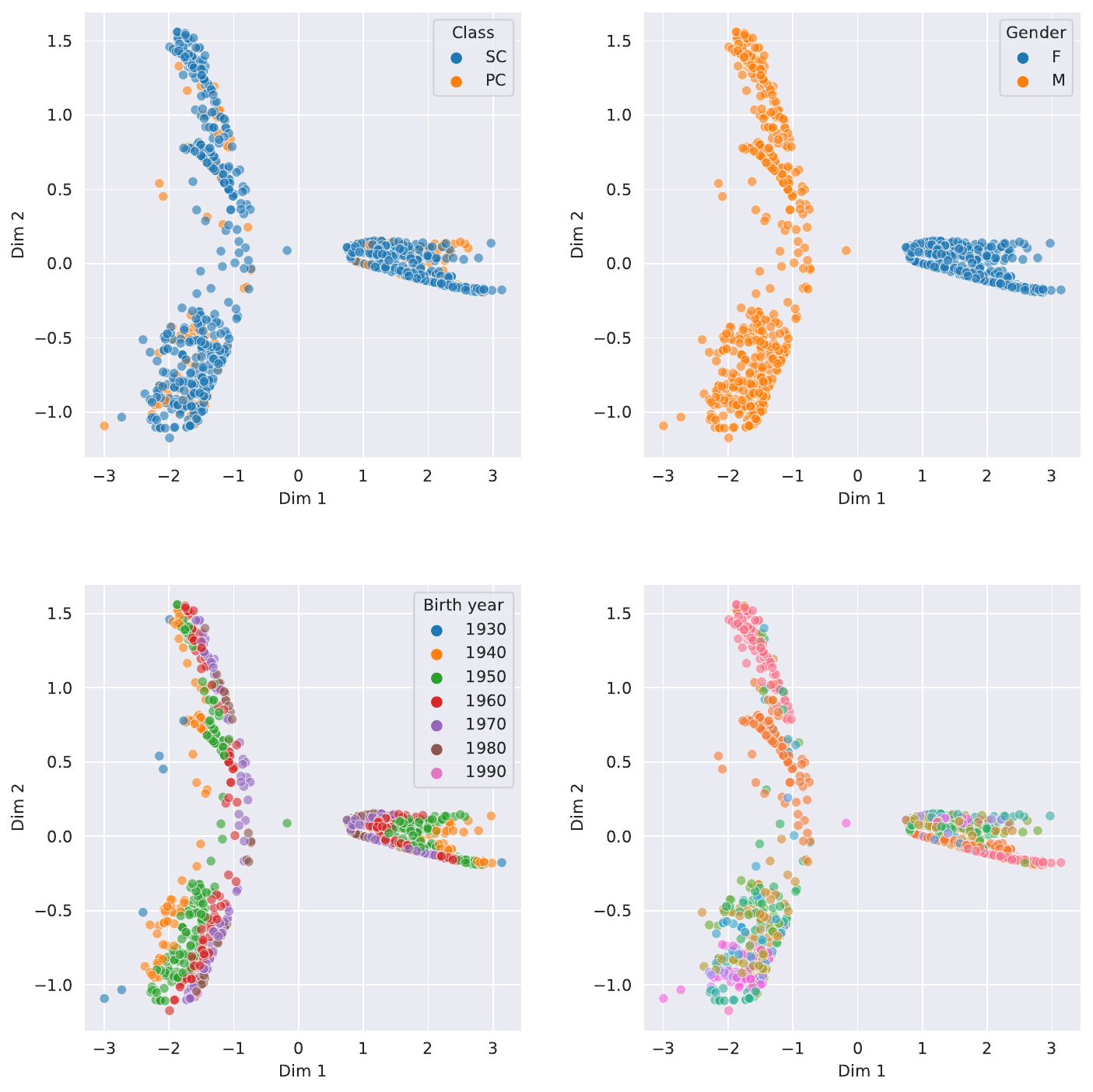}
   \caption{Node embeddings extracted from the Attri2Vec model with the social features reduced to two dimension using ISOMAP with 100 neighbours. In the top left plot, the node embedding separates primary care doctors (PC) from specialty doctors (SC); in the top right, the embeddings are colored by gender.The embeddings naturally show at least three major clusters of doctors, which clearly are well separated by gender. As expected, the distribution of primary care doctors vs specialist doctors is similar throughout all clusters. In the bottom left the node embedding is colored by birth decade of the doctors whereas in the right, the colors represent each hospital.There is a tendency of doctors who work in the same hospital to have a similar embedding --- this evidence showcases the importance of the social features in representing this information in the model.}
   \label{fig:isomap-with}
\end{figure}

%\paragraph*{S4 Fig.}
%\label{tsne with}
%{\bf  Node embeddings extracted from the Attri2Vec model with the social features reduced to two dimensions using T-SNE with perplexity = 40} In the top left plot, the node embedding separates primary care doctors (PC) from specialty doctors (SC); in the top right the embeddings are colored by gender. The embeddings naturally show some major clusters of doctors, which clearly are well separated by gender. As expected, the distribution of primary care doctors vs specialist doctors is similar throughout all clusters. The contrast of this embedding with the embedding of fig ~\ref{fig:tsne without\_1} highlights the increase on the embedding structures produced by the addition of the social features. In the bottom left, the node embeddings are colored by the birth year of the doctors, whereas in the bottom right, the colors represent each hospital. The doctors who work in the same hospital are mostly represented in the same cluster--- this evidences, once more, the importance of the social features in representing this information in the model and in predicting the referrals. The birth year is also stratified by cluster but is not sufficient to increase the performance of the referral predictor

\begin{figure}
    \centering
    \includegraphics[scale=0.5]{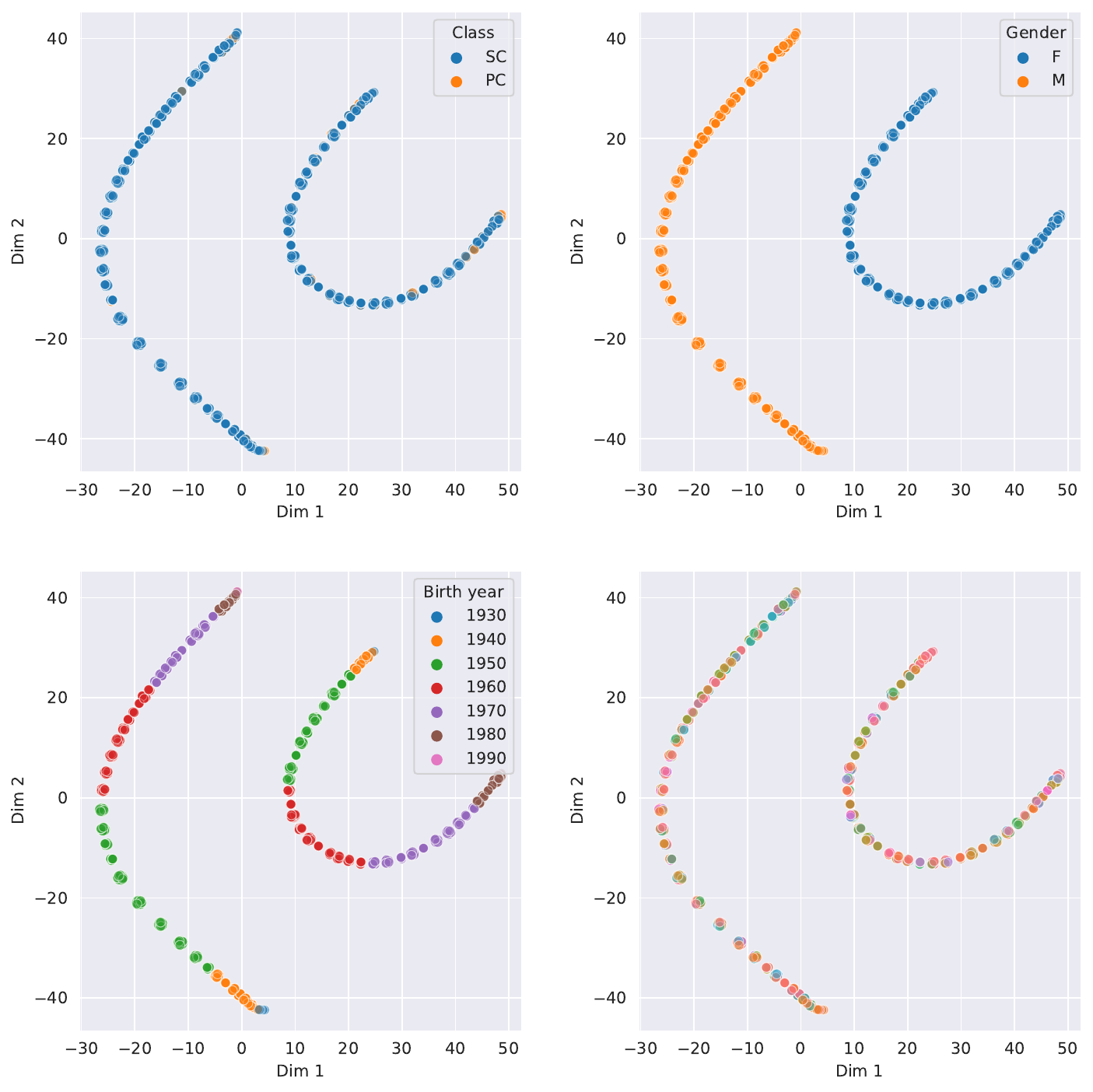}
    \caption{ Node embeddings extracted from the Attri2Vec model without the social features reduced to two dimension using T-SNE with perplexity = 40. In the top left the node embedding separates primary care doctors (PC) from specialty doctors (SC); in the top right the embeddings are colored by gender. The gender is the most important feature in the decision on the model without the social features --- the model haven't the power to decompose the embeddings further.  In the bottom left the node embeddings are colored by birth year of the doctors whereas in the right, the colors represent each hospital.The pattern present in the ISOMAP visualizations is also present here. Without the social features, the information about hospital/ region where the clinician work not contribute to the prediction.  }
    \label{fig:tsne-without-1}
\end{figure}

%\paragraph*{S5 Fig.}
%\label{isomap_with}
%{\bf Node embeddings extracted from the Attri2Vec model with the social features reduced to two dimensions using ISOMAP with 100 neighbors} In the top left plot, the node embedding separates primary care doctors (PC) from specialty doctors (SC); in the top right, the embeddings are colored by gender. The embeddings naturally show at least three major clusters of doctors, which clearly are well separated by gender. As expected, the distribution of primary care doctors vs specialist doctors is similar throughout all clusters. In the bottom left the node embedding is colored by birth decade of the doctors whereas in the right, the colors represent each hospital.There is a tendency of doctors who work in the same hospital to have a similar embedding --- this evidence showcases the importance of the social features in representing this information in the model.

\begin{figure}
    \centering
    \includegraphics[scale=0.5]{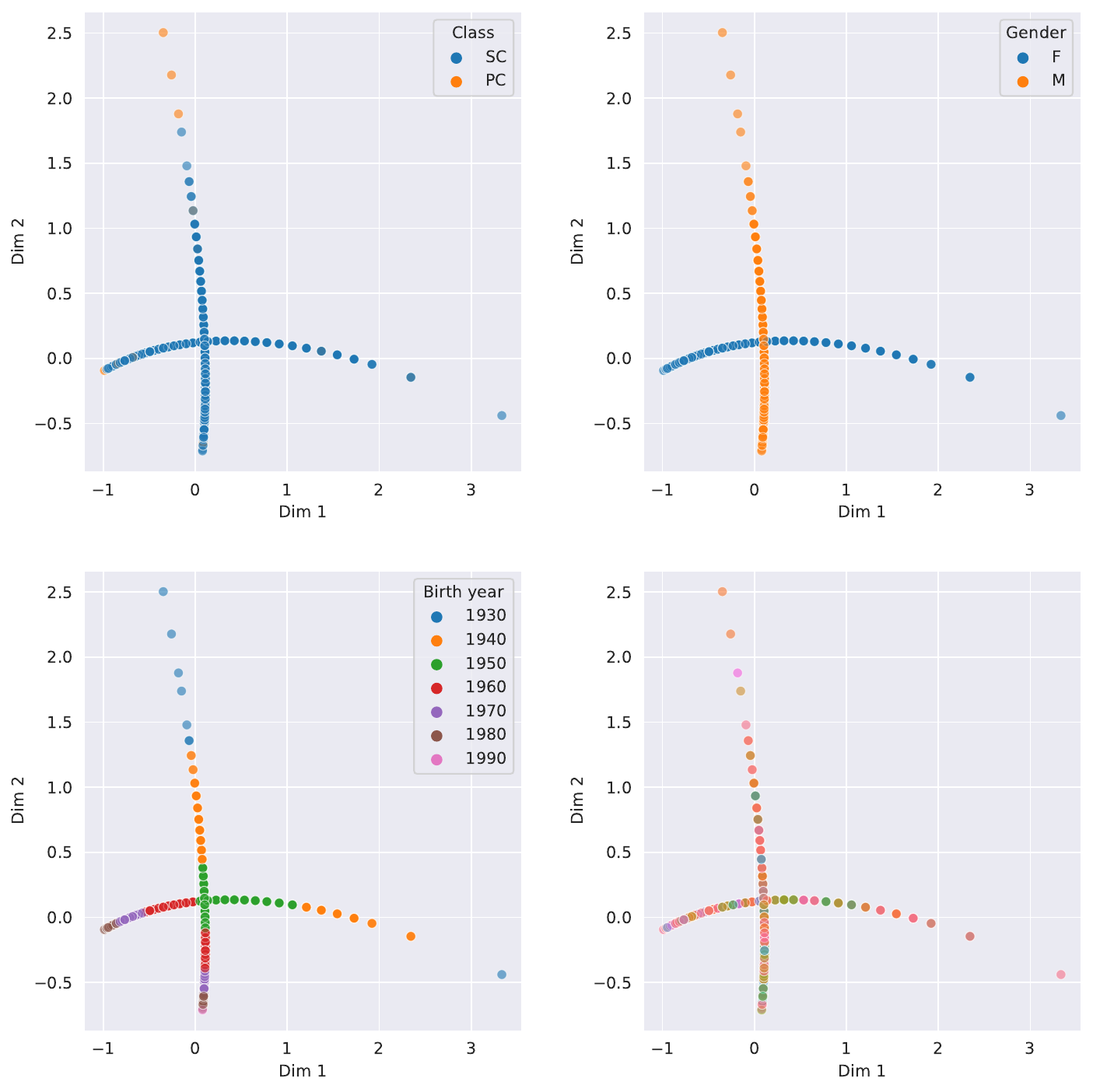}
   \caption{Node embeddings extracted from the Attri2Vec model without the social features reduced to two dimension using ISOMAP with 100 neighbours In the top left plot, the node embedding separates primary care doctors (PC) from specialty doctors (SC) whereas in the top right the embeddings are colored by gender. These embeddings show a poorer structure than the one without social features \ref{isomap_with} which evidences the importance of the social features. Gender is the main factor that can differentiate the embeddings. In the bottom plots, the node embedding is colored by birth decade of the doctors (left) and by hospitals where they work (right). Without the social information, the embeddings don't seem to provide any information about the hospital the doctors work in. In contrast, the age seems to be the second most important factor that give structure to the embeddings}
    \label{fig:isomap-without-1}
\end{figure}

\end{document}